\pdfoutput=1
\documentclass[11pt]{article}

\usepackage[final]{acl}

\usepackage{times}
\usepackage{latexsym}
\usepackage[T1]{fontenc}
\usepackage[utf8]{inputenc}
\usepackage{microtype}
\usepackage{inconsolata}

\usepackage{graphicx}
\usepackage{booktabs}
\usepackage{multirow}
\usepackage{amsmath}
\usepackage{tcolorbox}
\usepackage{adjustbox}
\usepackage{siunitx}
\usepackage{pgfplots}
\usepgfplotslibrary{groupplots}
\pgfplotsset{compat=1.18}

\title{CAF-Score: Calibrating CLAP with LALMs for Reference-free Audio-Text Alignment Evaluation}

\author{
  Insung Lee \quad
  Taeyoung Jeong \quad
  HaeJun Yoo \quad
  Du-Seong Chang\thanks{Co-corresponding authors.} \quad
  Myoung-Wan Koo\footnotemark[1] \\
  Sogang University \\
  \texttt{\{dlstjd6474, john428, judejiwoo, dschang, mwkoo\}@sogang.ac.kr}
}

\begin{document}
\maketitle

\begin{abstract}
While Large Audio-Language Models (LALMs) have advanced audio understanding and generation, robustly evaluating audio-text alignment remains difficult: reference-based metrics require costly annotations, while Contrastive Language-Audio Pretraining (CLAP)-based reference-free metrics frequently overlook syntactic errors and fine-grained details. We propose \textbf{CAF-Score}, a reference-free metric that calibrates CLAP's coarse-grained semantic alignment with the fine-grained comprehension and syntactic awareness of LALMs. Because CAF-Score scores an audio-text pair symmetrically, one configuration applies to both alignment directions: audio captioning evaluation (the BRACE benchmark) and text-to-audio (TTA) generation evaluation (the RELATE and PAM benchmarks), with a single fusion weight selected once on BRACE and transferred without re-tuning. Across all three human-annotated benchmarks, CAF-Score improves agreement with human judgments over standalone CLAP in the large majority of CLAP--LALM combinations---and in all combinations with the strongest LALM backbone---and outperforms standalone LALM metrics on the main captioning benchmark. It also remains competitive with reference-based metrics in challenging scenarios.\footnote{Code: \url{https://github.com/inseong00/CAF-Score}}
\end{abstract}

\section{Introduction}

Recent advances in Large Audio-Language Models (LALMs) have demonstrated strong capabilities in fine-grained audio understanding and reasoning~\citep{comanici2025gemini, goel2025audio, xu2025qwen3omni}. These developments have, in turn, spurred the emergence of specialized audio captioning models~\citep{xu2025qwen3omni, dinkel2025midashenglm} and increasingly capable text-to-audio (TTA) generation systems~\citep{liu2023audioldm, liu2024audioldm}. Both directions share a common evaluation bottleneck: measuring how well an audio signal and a piece of text align with each other, without access to costly ground-truth references.

\begin{figure}[t]
    \centering
    \includegraphics[width=0.9\columnwidth]{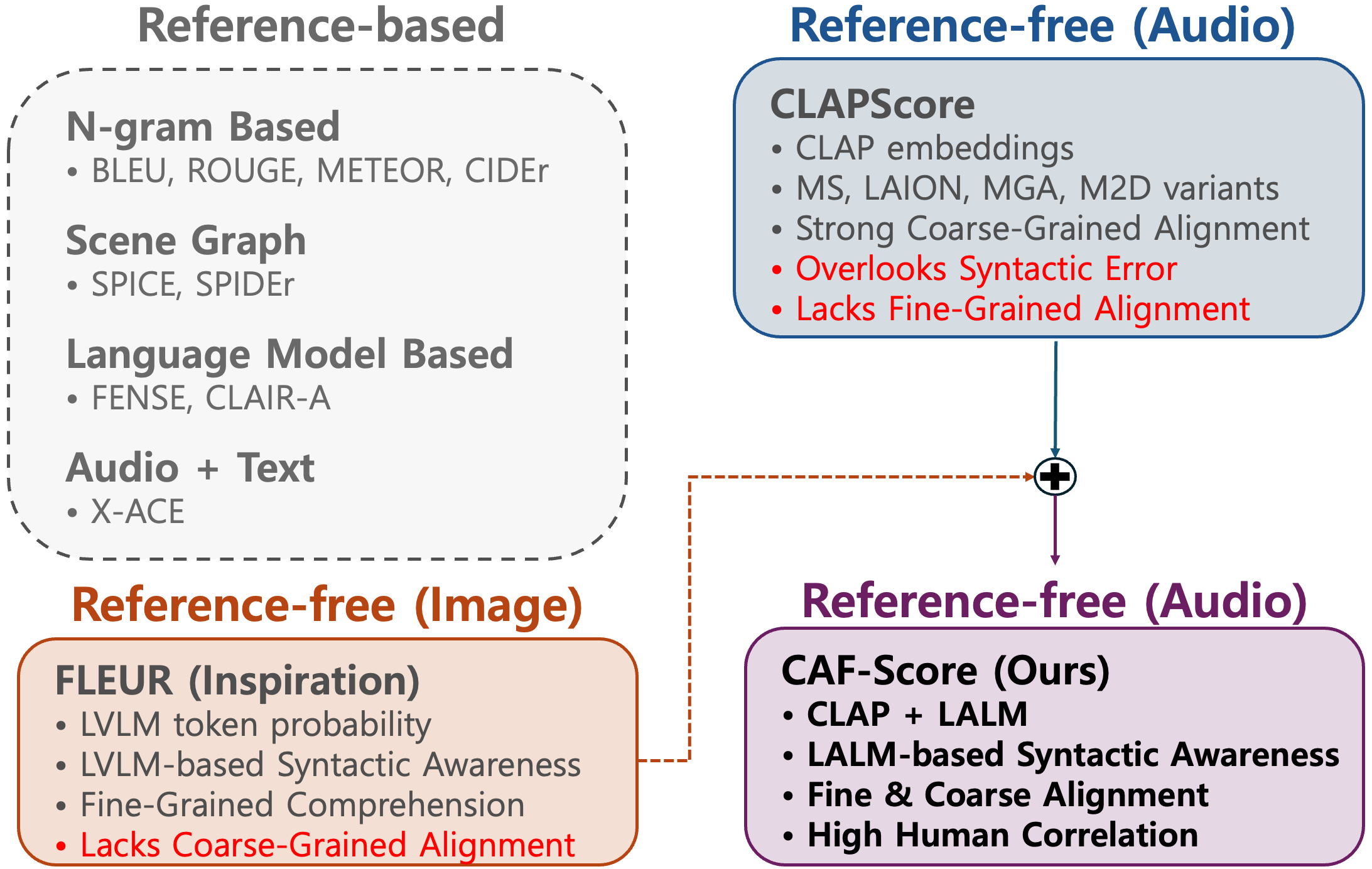}
    \caption{\textbf{Overview of audio-text alignment evaluation metrics.} Reference-based metrics (left) depend on ground-truth captions, while CLAP-based reference-free metrics (top right) lack syntactic and fine-grained sensitivity. \textbf{CAF-Score} integrates CLAP's coarse-grained alignment with the fine-grained reasoning of LALMs, inspired by FLEUR~\citep{lee2024fleur} from the vision domain.}
    \label{fig:caf_introduce}
\end{figure}

As illustrated in Figure~\ref{fig:caf_introduce}, most existing metrics rely on text-to-text comparison against ground-truth references---from n-gram and scene-graph metrics~\citep{papineni2002bleu, lin2004rouge, banerjee2005meteor, vedantam2015cider, anderson2016spice, liu2017improved} to embedding- and LLM-based approaches such as FENSE~\citep{zhou2022can} and CLAIR-A~\citep{wu2024clair}---and even audio-aware variants like X-ACE~\citep{wang2024x} still necessitate reference captions, limiting scalability. Reference-free alternatives inspired by vision-domain paradigms~\citep{hessel2021clipscore, lee2024fleur} have emerged in the audio domain in the form of CLAP-based metrics~\citep{elizalde2023clap, wu2023large, li2024advancing, niizumi2025m2d}; however, these often fail to capture syntactic errors and fine-grained semantic distinctions due to their predominantly coarse-grained contrastive training objectives~\citep{guo2025brace}.

To overcome this limitation, we propose the \textbf{CLAP-aligned FLEUR score (CAF-Score)}. By calibrating CLAPScore using FLEUR~\citep{lee2024fleur}, CAF-Score combines the robust coarse-grained semantic alignment of CLAP models with the syntactic sensitivity and fine-grained semantic reasoning capabilities of LALMs. Because both components score a single audio-text pair symmetrically, the same framework applies to both directions of the alignment problem: judging a caption against a given audio (audio captioning evaluation) and judging a generated audio against a given text prompt (TTA evaluation). This direction-agnostic design distinguishes CAF-Score from the concurrent AQAScore~\citep{kuan2026aqascore}, an LALM-based metric highly effective for TTA evaluation but, by its authors' own account, only CLAPScore-level in caption selection. Our hybrid approach improves agreement with human judgments over standalone CLAP-based metrics in the large majority of evaluated backbone combinations---and in all combinations with the strongest LALM backbone---and over standalone LALM-based metrics in the main evaluation settings.

\begin{figure*}[t]
    \centering
    \includegraphics[width=1.0\linewidth]{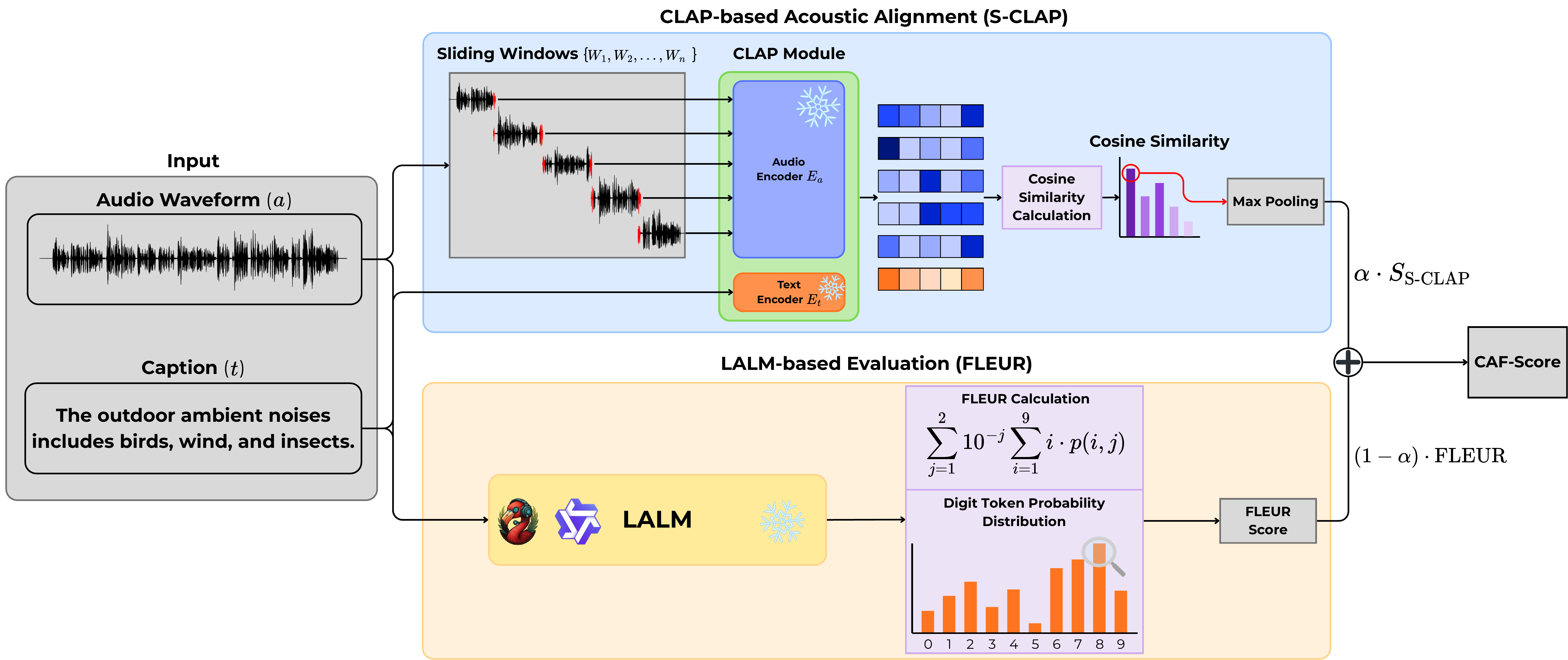}
    \caption{Overall architecture of CAF-Score. The CLAP branch applies a sliding-window strategy with Max pooling to obtain the S-CLAPScore; the LALM branch computes a FLEUR score from the probability distribution over digit tokens. The final CAF-Score is a weighted combination of the two, computed entirely at inference time with frozen backbones.}
    \label{fig:pipeline}
\end{figure*}

The contributions of this study are threefold:
\begin{itemize}
    \item We extend the FLEUR methodology to the audio domain, demonstrating the effectiveness of LALMs for evaluating audio-text alignment.
    \item We introduce the \textbf{CAF-Score}, a novel reference-free metric that leverages LALMs to capture syntactic structure and fine-grained semantic nuances overlooked by CLAP-based approaches, achieving competitive performance with even reference-based baselines.
    \item We validate the framework in both alignment directions---audio captioning evaluation on BRACE, and TTA evaluation on RELATE and PAM---across multiple CLAP backbones and state-of-the-art open-source LALMs, using a single direction-agnostic configuration: one shared prompt formulation and one fusion weight selected once on BRACE, transferred without re-tuning.
\end{itemize}

\section{Related Work}
\label{sec:related_work}
\subsection{Contrastive Audio-Text Alignment}
\label{sec:related_clap}
To address the need for reference-free evaluation, contrastive learning-based metrics have gained prominence. Building on the success of CLIPScore~\citep{hessel2021clipscore} in the vision domain, CLAPScore~\citep{elizalde2023clap} computes cosine similarity between audio and text embeddings within a shared latent space. Several CLAP variants have since been proposed to improve audio-text alignment, including LAION-CLAP~\citep{wu2023large} for variable-length audio processing, MGA-CLAP~\citep{li2024advancing} for fine-grained frame-level alignment, and M2D-CLAP~\citep{niizumi2025m2d} for general-purpose representation learning. Despite their overall robustness, results from the BRACE benchmark~\citep{guo2025brace} indicate that these embedding-based metrics frequently fail to capture fine-grained acoustic characteristics and syntactic errors, largely due to the inherently coarse-grained nature of vector-space alignment.

\subsection{Generative Evaluation with LALMs}
\label{sec:related_lalm}
In the vision domain, FLEUR~\citep{lee2024fleur} has shown that Large Vision-Language Models (LVLMs) can effectively assess vision-language alignment through smoothed token probabilities. Robust progress in LALMs has likewise demonstrated strong capabilities in fine-grained audio understanding and syntactic awareness~\citep{comanici2025gemini, goel2025audio, xu2025qwen3omni}, making them natural candidates for generative evaluation in the audio domain. However, when used as standalone evaluators, LALMs remain subject to well-documented limitations, including hallucinations~\citep{hsu2025reducing,guo2025brace}, which undermine their reliability for consistent evaluation.

The work closest to ours is the concurrent AQAScore~\citep{kuan2026aqascore}, which reformulates TTA evaluation as probabilistic semantic verification: an LALM is asked a yes/no question about whether the audio contains the sound events described by the text, and the normalized probability of a ``Yes'' answer serves as the alignment score. When extended to caption evaluation on BRACE via a comparative yes/no question posed over each caption pair, its authors report accuracy comparable to CLAPScore, with notably lower accuracy on human--machine (HM) pairs; they attribute this to a judging-preference mismatch---the verification formulation favors perceptually detailed, event-by-event captions, whereas human annotators often prefer concise semantic summaries---and identify such inconsistency in text-selection settings as a limitation of their framework. Our work differs in two respects. First, we adapt FLEUR's grading formulation---a continuous score computed from the probability distribution over digit score tokens, previously unexplored in the audio-language domain---which elicits a graded quality judgment rather than a binary containment verification, and we apply the same prompt formulation (perspective-inverted) in both alignment directions. Second, rather than relying on the LALM alone, CAF-Score anchors the LALM's fine-grained judgment to CLAP's coarse-grained acoustic alignment, a grounding mechanism designed to temper exactly the kind of caption-level bias reported above. This hybrid design enables more reliable reference-free audio-text evaluation across both directions with a single configuration.

\section{Methodology}
\label{sec:method}

We formulate reference-free evaluation as scoring the alignment $s(x_a, x_t)$ between an audio signal $x_a$ and a text $x_t$. This single formulation instantiates both tasks considered in this work: in audio captioning evaluation, the audio is given and a candidate caption is assessed; in TTA evaluation, the text prompt is given and a generated audio is assessed. As illustrated in Figure~\ref{fig:pipeline}, CAF-Score combines two independent components computed on the same pair: an S-CLAPScore from CLAP and a FLEUR score from an LALM.

\subsection{CLAP-based Alignment (S-CLAPScore)}

CLAP models comprise a text encoder $E_t$ and an audio encoder $E_a$. Given a candidate text $x_t$ and an audio input $x_a$, the standard CLAPScore $S_\text{{CLAP}}$ is computed as the cosine similarity between their L2-normalized embeddings:
\begin{equation}
    S_\text{{CLAP}}(x_a, x_t) = \text{Norm}\big(E_t(x_t)\big)^\top \text{Norm}\big(E_a(x_a)\big)
\end{equation}
where $\text{Norm}(\cdot)$ denotes L2-normalization.

Standard CLAP models are constrained by a fixed maximum audio input duration, so naive truncation of variable-length audio can discard critical acoustic information~\citep{li2024advancing}. We therefore adopt a sliding-window strategy following the BRACE evaluation protocol~\citep{guo2025brace}, which averages window-level scores. However, this averaging causes semantic dilution: captions often refer to brief but salient events occupying only a small portion of the audio~\citep{li2024advancing}, whose high alignment score is attenuated by unrelated background or silent segments. We instead adopt Max pooling to identify the best-matching segment, and denote the resulting metric S-CLAPScore (Sliding-window CLAPScore) $S_\text{{S-CLAP}}$:
\begin{equation}
    S_\text{{S-CLAP}}(x_a, x_t) = \max_{n=1}^{N} \big( S_\text{{CLAP}}(x_{w_n}, x_t) \big)
\end{equation}
where $\{x_{w_n}\}_{n=1}^N$ denotes the set of sliding-windows extracted from the audio $x_a$. A hop size of \SI{1}{s} is used for all models, and the window length is set to each model's maximum supported input duration (\SI{7}{s} for MS-CLAP, \SI{10}{s} for the other CLAP variants).

\subsection{LALM-based Evaluation (FLEUR)}
Although LALMs can directly produce discrete numerical scores (e.g., ``0.85''), we observed that such raw outputs frequently tie among candidates, obscuring subtle quality differences.

To overcome this limitation, we adopt FLEUR~\citep{lee2024fleur}, which computes a continuous score from the token-level probability distribution over digit tokens rather than from the deterministic numeric output. We prompt the LALM with a grading task adapted to the target domain by replacing ``image'' with ``audio'' in the original prompt from~\citet{lee2024fleur}; the exact templates for both settings are provided in Appendix~\ref{sec:appendix_prompts}. Specifically, let $p(i, j)$ denote the probability of digit token $i \in \{1, \dots, 9\}$ appearing at the $j$-th decimal place. The FLEUR score is defined as:
\begin{equation}
    \text{FLEUR} = \sum_{j=1}^{2} 10^{-j} \sum_{i=1}^{9} i \cdot p(i, j)
\end{equation}

Intuitively, two captions that both produce ``0.85'' as raw output can have different underlying probability distributions, yielding distinct FLEUR scores (e.g., 0.876 vs.\ 0.851) and thereby resolving ties.

\subsection{CAF-Score (CLAP-aligned FLEUR score)}

To combine CLAP's robust coarse-grained alignment with the fine-grained semantic and syntactic sensitivity of LALMs---while tempering the weaknesses of each component discussed in Section~\ref{sec:related_work}---we propose the \textbf{CAF-Score (CLAP-aligned FLEUR score)}: S-CLAPScore serves as the foundation for coarse-grained semantic alignment, and FLEUR calibrates it with fine-grained semantic and syntactic assessment.

Here, calibrates denotes score-level fusion: as defined below, the FLEUR score adjusts the S-CLAPScore through a fixed linear combination, rather than referring to probability calibration or a learned calibration model.

The final CAF-Score is computed as a linear combination of the two component scores using a weighting parameter $\alpha$:
\begin{equation}
    S_{\text{CAF}} = \alpha \cdot S_{\text{S-CLAP}} + (1 - \alpha) \cdot \text{FLEUR}
\end{equation}
where $\alpha$ controls the trade-off between the two components. We set $\alpha=0.8$ globally: although the per-backbone optimum varies, this value is selected once on BRACE-Main by average performance across backbones. It is kept fixed for all other evaluations---including the TTA benchmarks in Section~\ref{sec:tta}---demonstrating that CAF-Score requires no per-configuration tuning. A sample-level adaptive weighting we explored did not outperform this fixed value on average (see Appendix~\ref{sec:appendix_alpha} for the per-backbone analysis, the adaptive-weighting comparison, and a discussion of score scales).

\section{Experiments}

\begin{table*}[t]
\centering
\caption{Accuracy on the BRACE benchmark. The upper section reports
reference-free metrics (standalone CLAP, standalone LALM, and the proposed
CAF-Score); the lower section reports reference-based baselines. Overall
denotes the sample-count-weighted average across subsets. AF3 denotes
AudioFlamingo3; within the CAF-Score rows, AF3 and Qwen3 correspond to
AudioFlamingo3-Base and Qwen3-Omni-Instruct, respectively.}
\label{tab:main_table}
\resizebox{\textwidth}{!}{%
\begin{tabular}{lccccccccc|ccc}
\toprule
\multirow{3}{*}{\textbf{Model (Method)}} & \multicolumn{9}{c}{\textbf{BRACE-Main}} & \multicolumn{3}{c}{\textbf{BRACE-Hallucination}}\\

\cmidrule(lr){2-10} \cmidrule(lr){11-13}
& \multicolumn{4}{c}{\textbf{AudioCaps-Main}} & \multicolumn{4}{c}{\textbf{Clotho-Main}} & \multirow{2}{*}{\textbf{Overall}} & \multirow{2}{*}{\textbf{AudioCaps-Hallu}} & \multirow{2}{*}{\textbf{Clotho-Hallu}} &  \multirow{2}{*}{\textbf{Overall}}\\
\cmidrule(lr){2-5} \cmidrule(lr){6-9}
 & HH & HM & MM & Total & HH & HM & MM & Total & & & \\
\midrule
\midrule
\multicolumn{13}{c}{\textbf{Reference-free}} \\
\midrule
\midrule
\multicolumn{13}{c}{\textit{CLAPScore (CLAP-ONLY)}} \\
\midrule

MS-CLAP & 64.03 & 48.76 & 55.08 & 53.71 & 67.07 & 70.92 & 65.93 & 67.95 & 61.42 & 72.99 & 85.23 & 80.61 \\
LAION-CLAP & 63.31 & 77.30 & 65.42 & 69.78 & 59.28 & 75.64 & 65.63 & 68.62 & 69.15 & 87.03 & 84.31 & 85.34 \\
MGA-CLAP & 62.59 & 73.48 & 65.24 & 68.12 & 71.26 & 67.58 & 63.41 & 65.95 & 66.95 & 92.18 & 87.26 & 89.12 \\
M2D-CLAP & \underline{65.47} & 80.22 & 63.99 & 70.48 & 67.66 & 66.21 & 62.81 & 64.69 & 67.35 & 91.03 & 83.93 & 86.61 \\
\midrule
\multicolumn{13}{c}{\textit{S-CLAPScore (CLAP-ONLY)}} \\
\midrule

MS-CLAP & 64.03 & 46.07 & 53.65 & 51.97 & 65.87 & 72.89 & 67.26 & 69.21 & 61.30 & 73.73 & 87.07 & 82.04 \\
LAION-CLAP & 63.31 & 77.30 & 65.42 & 69.78 & 59.88 & 75.83 & 66.96 & 69.43 & 69.59 & 87.03 & 84.45 & 85.42 \\
MGA-CLAP & 62.59 & 73.48 & 65.24 & 68.12 & \underline{72.46} & 68.96 & 62.52 & 66.17 & 67.06 & 92.18 & 88.08 & 89.63 \\
M2D-CLAP & \underline{65.47} & 80.22 & 63.99 & 70.48 & 70.06 & 69.94 & 63.85 & 66.91 & 68.55 & 91.03 & 84.47 & 86.95 \\
\midrule
\multicolumn{13}{c}{\textit{Raw Score (LALM-ONLY)}} \\
\midrule

AF3-Base & 33.81 & 28.09 & 35.12 & 32.23 & 13.77 & 22.40 & 23.26 & 21.76 & 26.56 & 25.28 & 60.83 & 47.41 \\
AF3-Think & 15.83 & 22.47 & 19.07 & 20.00 & 21.56 & 21.22 & 21.33 & 21.32 & 20.71 & 24.81 & 21.90 & 23.00 \\
Qwen3-Omni-Instruct & 16.55 & 48.54 & 41.00 & 40.96 & 35.93 & 54.62 & 45.04 & 47.52 & 44.51 & 81.78 & 82.47 & 82.21 \\
Qwen3-Omni-Thinking & 25.90 & 55.73 & 43.14 & 45.94 & 41.92 & 55.40 & 42.22 & 47.15 & 46.59 & 81.86 & 75.80 & 78.09 \\
\midrule
\multicolumn{13}{c}{\textit{FLEUR (LALM-ONLY)}} \\
\midrule

AF3-Base & 53.24 & 52.58 & 66.49 & 59.48 & 63.47 & 68.37 & 61.19 & 64.17 & 62.02 & 93.83 & 95.86 & 95.09 \\
AF3-Think & 41.73 & 52.58 & 56.51 & 53.19 & 59.28 & 54.81 & 55.11 & 55.11 & 54.45 & 85.80 & 90.14 & 88.50 \\
Qwen3-Omni-Instruct & 53.24 & 78.65 & 66.31 & 69.52 & 67.66 & 74.66 & 61.63 & 67.28 & 68.31 & \textbf{98.48} & \textbf{98.00} & \textbf{98.18} \\
Qwen3-Omni-Thinking & 51.80 & 71.46 & 58.82 & 62.88 & 58.68 & 68.96 & 56.74 & 61.58 & 62.18  & 89.70 & 86.99 & 88.01 \\
\midrule
\multicolumn{13}{c}{\textit{CAF-Score (Ours, CLAP + LALM)}} \\
\midrule

AF3+MS-CLAP & 62.59 & 48.09 & 60.07 & 55.72 & 64.67 & 73.87 & \underline{68.74} & 70.17 & 63.54 & 88.16 & 92.98 & 91.16 \\
AF3+LAION-CLAP & 63.31 & 76.63 & 67.38 & 70.48 & 58.68 & 77.01 & 67.85 & 70.17 & 70.31 & 93.20 & 89.98 & 91.20 \\
AF3+MGA-CLAP & \underline{65.47} & 71.01 & 67.91 & 68.82 & 70.06 & 70.92 & 63.26 & 66.99 & 67.83 & 95.84 & 93.20 & 94.20 \\
AF3+M2D-CLAP & 64.03 & 70.34 & 67.38 & 68.12 & 67.66 & 71.71 & 63.70 & 67.21 & 67.63 & 95.69 & 92.38 & 93.63 \\
Qwen3+MS-CLAP & 64.03 & 60.67 & 63.99 & 62.71 & 67.66 & \underline{79.17} & \textbf{70.07} & \textbf{73.21} & 68.39 & 93.52 & 96.61 & 95.44 \\
Qwen3+LAION-CLAP & 63.31 & \underline{81.57} & \textbf{68.09} & \underline{72.75} & 61.68 & \textbf{80.16} & 68.00 & \underline{71.80} & \underline{72.24} & 96.24 & 94.94 & 95.43 \\
Qwen3+MGA-CLAP & 63.31 & 81.12 & 67.74 & 72.40 & \textbf{73.65} & 75.44 & 63.85 & 69.43 & 70.79 & 97.57 & \underline{96.80} & 97.09 \\
Qwen3+M2D-CLAP & \textbf{67.63} & \textbf{86.97} & 67.91 & \textbf{75.28} & 70.66 & 77.80 & 66.52 & 71.28 & \textbf{73.11} & \underline{97.96} & 96.59 & \underline{97.11} \\
\midrule
\midrule
\multicolumn{13}{c}{\textbf{Reference-based}} \\
\midrule
\midrule
FENSE~\citep{zhou2022can} & 61.15 & 84.49 & 67.38 & 73.28 & \textbf{56.89} & 84.68 & 64.15 & 70.98 & 72.04 & \textbf{96.76} & \textbf{96.18} & \textbf{96.40} \\
CLAIR-A~\citep{wu2024clair} & \textbf{66.19} & \textbf{90.79} & \textbf{67.74} & \textbf{76.51} & 56.29 & \textbf{91.94} & \textbf{65.63} & \textbf{74.39} & \textbf{75.36} & 91.45 & 91.55 & 91.51\\
\bottomrule
\end{tabular}%
}
\end{table*}

\subsection{Benchmarks}
\label{sec:benchmarks}

\paragraph{Audio Captioning Evaluation.}
We use BRACE~\citep{guo2025brace}, a reference-free benchmark for audio captioning assessment built on the evaluation splits of AudioCaps~\citep{kim-etal-2019-audiocaps} and Clotho~\citep{drossos2019clothoaudiocaptioningdataset}. It comprises BRACE-Main (2,496 caption pairs for fine-grained caption comparison) and BRACE-Hallucination (2,027 audio clips paired with noun-modified captions). Unlike existing alternatives---the FENSE benchmark~\citep{zhou2022can}, built on earlier-generation captioning systems, and hallucination-focused benchmarks such as Comp-A~\citep{ghosh2023compa} and AHa-Bench~\citep{cheng2025ahabench}, which remain reference-based---BRACE is designed for modern LALMs and supports reference-free evaluation of both misalignment and hallucination. We depart from the original pairwise protocol and score each caption independently against its audio, which enables caption-level analysis while still permitting reconstruction of the original pairwise preferences (Appendix~\ref{sec:appendix_brace}).

\vspace{-3mm}
\paragraph{Text-to-Audio Generation Evaluation.}
To validate the opposite alignment direction, we use RELATE~\citep{kanamori2025relate} and PAM~\citep{deshmukh2024pam}, which provide gold-standard human subjective ratings of how well generated audio matches its text prompt. Following standard practice, we report Pearson (LCC), Spearman (SRCC), and Kendall's Tau (KTAU) correlations between metric scores and human ratings.

\subsection{Experimental Setup}
\label{sec:experimental_setup}
\vspace{-2mm}
We evaluate four representative CLAP backbones (MS-CLAP, LAION-CLAP, MGA-CLAP, and M2D-CLAP) and two LALM families (AudioFlamingo3 and Qwen3-Omni), covering both standard and reasoning variants. On the TTA benchmarks, we additionally evaluate the Qwen-2.5-Omni~\citep{wei2025qwen25omni} family as a backbone-controlled comparison with AQAScore~\citep{kuan2026aqascore}, which is built on the same backbone. AQAScore also reports results on BRACE, obtained by posing a comparative yes/no question jointly over each caption pair; since this protocol differs from our independent single-caption scoring (Appendix~\ref{sec:appendix_brace}) and its implementation is not publicly available at the time of writing, we cite their self-reported numbers in Section~\ref{sec:brace_main} rather than including them in Table~\ref{tab:main_table}. Full model specifications, implementation details, and the component-wise evaluation protocol are provided in Appendix~\ref{sec:appendix_setup}.

\paragraph{Benchmark statistics.}
BRACE-Main contains 2,496 caption pairs: AudioCaps-Main comprises 1,145 pairs over 275 clips (HH/HM/MM: 139/445/561), and Clotho-Main comprises 1,351 pairs over 300 clips (167/509/675). BRACE-Hallucination contains 2,027 clips (765 from AudioCaps and 1,262 from Clotho), each paired with five original--hallucinated caption pairs, yielding 3,825 and 6,310 pairs, respectively. Since multiple pairs share the same clip, pairs within a subset are not fully independent; we therefore interpret small accuracy differences, particularly on the HH subsets, with caution.

\paragraph{Computational cost.}
CAF-Score is inference-only with frozen backbones and requires no training. For each audio--text pair, it requires CLAP encoder passes (one per sliding window) and a single LALM generation of a short numeric score; FLEUR is computed from the token probabilities returned by this same generation, so smoothing requires no additional LALM calls. The LALM call dominates the cost: with Qwen3-Omni-30B-A3B served by vLLM on a single NVIDIA A100 (80GB), the measured latency is approximately \SI{0.28}{s} per LALM call at batch size 1. This is considerably more expensive than CLAP-only metrics, which require only encoder forward passes, but remains practical for offline evaluation.

\section{Results}
\label{sec:results}
\subsection{Audio Captioning Evaluation}
\label{sec:captioning_results}

We first evaluate the direction in which the audio is given and candidate captions are assessed, using the two subsets of BRACE.

\subsubsection{BRACE-Main}
\label{sec:brace_main}

Quantitative results on the BRACE-Main benchmark are summarized in Table~\ref{tab:main_table}. All reported scores represent accuracy against human preference judgments, so higher accuracy directly reflects stronger correlation with human evaluation. Two component-level trends, analyzed in detail in Appendix~\ref{sec:appendix_design}, are visible in the table: \textbf{S-CLAPScore} consistently outperforms the truncated CLAPScore baseline on Clotho-Main (where audio is long enough for windowing to matter for all backbones), and the probabilistic \textbf{FLEUR} metric substantially outperforms raw discrete LALM scores, whose frequent ties reduce discriminative power.

Comparing reasoning and non-reasoning variants, the non-reasoning variants performed better under FLEUR for both families---AudioFlamingo3 in all settings, and Qwen3-Omni despite its reasoning variant scoring higher on raw outputs. We therefore use the \textbf{non-reasoning variants} (AudioFlamingo3-Base and Qwen3-Omni-Instruct) for the final CAF-Score.

The proposed \textbf{CAF-Score}, which integrates both CLAP and LALM components, achieves the highest reference-free performance on both AudioCaps-Main and Clotho-Main subsets. Notably, combining CLAP with \textbf{Qwen3-Omni-Instruct} outperforms configurations using AudioFlamingo3, likely reflecting the difference in training scale between the two families; \citet{kuan2026aqascore} report an analogous gap between Qwen-2.5-Omni and AudioFlamingo3. Specifically, the M2D-CLAP + Qwen3-Omni-Instruct combination achieves the best performance on AudioCaps-Main and overall, while MS-CLAP + Qwen3-Omni-Instruct attains the highest performance on Clotho-Main.

To contextualize these results, we additionally compare CAF-Score against two representative reference-based metrics: FENSE~\citep{zhou2022can}, which combines Sentence-BERT embeddings with a grammar error detector, and CLAIR-A~\citep{wu2024clair}, which leverages GPT-4o for reasoning-based semantic distance calculation. Despite operating without reference captions, CAF-Score achieves competitive overall performance against these baselines. Notably, CAF-Score remains competitive in the more challenging HH (Human-Human) scenarios, where both candidate captions are written by humans and exhibit subtle quality differences: its best configurations match or exceed the reference-based baselines on both AudioCaps and Clotho. However, the HH subsets are small (139 and 167 pairs), and on AudioCaps the margin over CLAIR-A amounts to only two pairs; we therefore regard CAF-Score as competitive with, rather than superior to, reference-based metrics in this regime.

Finally, we situate CAF-Score against the concurrent AQAScore~\citep{kuan2026aqascore}, which reports BRACE results under a comparative pairwise-question protocol: its best configurations reach 69.5 on AudioCaps-Main and 69.1 on Clotho-Main---comparable to CLAPScore, as its authors note---whereas the best CAF-Score configurations reach 75.28 and 73.21 under our protocol. While the protocol difference precludes an exact comparison, a backbone-controlled view is informative: with the same AudioFlamingo3 backbone, AQAScore reports 60.4 and 60.1 on the two subsets, whereas AF3-based CAF-Score reaches 70.48 and 70.17. 
\citet{kuan2026aqascore} trace their weakest results to the HM subsets, attributing them to a preference for exhaustive event-by-event captions over the concise summaries favored by human annotators. We note that their comparative questioning is precisely the style of pairwise prompting our single-caption protocol is designed to avoid (Appendix~\ref{sec:appendix_brace}). Moreover, grounding the LALM judgment in CLAP's acoustic alignment is the mechanism by which CAF-Score is designed to temper such caption-level biases.

\begin{table*}[t!]
\centering
\renewcommand{\arraystretch}{0.75}
\caption{Full performance comparison on TTA benchmarks including RELATE and PAM datasets. We report Pearson (LCC), Spearman (SRCC), and Kendall's Tau (KTAU) correlations with human judgments. All CAF-Score results use $\alpha=0.8$, transferred from BRACE without re-tuning. RELATE and AQAScore rows are taken from the original papers.}
\label{tab:tta_full_results}
\footnotesize
\begin{tabular}{lccc|ccc}
\toprule
\multirow{2}{*}{\textbf{Model}} & \multicolumn{3}{c}{\textbf{RELATE}} & \multicolumn{3}{c}{\textbf{PAM}} \\
\cmidrule(lr){2-4} \cmidrule(lr){5-7}
& LCC & SRCC & KTAU & LCC & SRCC & KTAU \\
\midrule
\midrule
RELATE~\citep{kanamori2025relate} & 0.385 & 0.383 & 0.265 & - & - & - \\
RELATE w/ CB~\citep{kanamori2025relate} & 0.377 & 0.374 & 0.259 & - & - & - \\
\midrule
\multicolumn{7}{c}{\textit{CLAPScore (CLAP-ONLY)}} \\
\midrule
MS-CLAP & 0.256 & 0.224 & 0.153 & 0.345 & 0.312 & 0.215 \\
LAION-CLAP & 0.404 & 0.386 & 0.268 & 0.461 & 0.452 & 0.314 \\
MGA-CLAP & 0.446 & 0.411 & 0.286 & 0.532 & 0.522 & 0.368 \\
M2D-CLAP & 0.492 & 0.457 & 0.322 & 0.566 & 0.548 & 0.388 \\
\midrule
\multicolumn{7}{c}{\textit{AQAScore~\citep{kuan2026aqascore} (LALM-ONLY)}} \\
\midrule
Qwen-2.5-Omni-3B & 0.443 & 0.453 & 0.327 & 0.540 & 0.560 & 0.410 \\
Qwen-2.5-Omni-7B & \textbf{0.544} & \textbf{0.556} & \textbf{0.396} & 0.518 & 0.589 & 0.429 \\
AF3-base & 0.475 & 0.508 & 0.357 & 0.496 & 0.538 & 0.383 \\
AF3-think & 0.435 & 0.474 & 0.330 & 0.582 & 0.587 & 0.419 \\
\midrule
\multicolumn{7}{c}{\textit{Raw Score (LALM-ONLY)}} \\
\midrule
Qwen-2.5-Omni-3B & 0.337 & 0.341 & 0.272 & 0.376 & 0.364 & 0.287 \\
Qwen-2.5-Omni-7B & 0.328 & 0.316 & 0.262 & 0.305 & 0.299 & 0.302 \\
AF3-Base & 0.120 & 0.111 & 0.092 & 0.126 & 0.089 & 0.132 \\
AF3-Think & 0.149 & 0.145 & 0.116 & -0.070 & -0.062 & -0.049 \\
Qwen3-Omni-Instruct & 0.317 & 0.348 & 0.282 & 0.420 & 0.384 & 0.299 \\
Qwen3-Omni-Thinking & 0.289 & 0.279 & 0.218 & 0.369 & 0.362 & 0.276 \\
\midrule
\multicolumn{7}{c}{\textit{FLEUR (LALM-ONLY)}} \\
\midrule
Qwen-2.5-Omni-3B & 0.363 & 0.408 & 0.287 & 0.372 & 0.411 & 0.288 \\
Qwen-2.5-Omni-7B & 0.399 & 0.407 & 0.286 & 0.418 & 0.429 & 0.302 \\
AF3-Base & 0.151 & 0.216 & 0.148 & 0.152 & 0.194 & 0.132 \\
AF3-Think & 0.163 & 0.163 & 0.111 & -0.025 & -0.042 & -0.029 \\
Qwen3-Omni-Instruct & 0.396 & 0.451 & 0.316 & 0.458 & 0.426 & 0.297 \\
Qwen3-Omni-Thinking & 0.292 & 0.246 & 0.171 & 0.367 & 0.351 & 0.242 \\
\midrule
\multicolumn{7}{c}{\textit{Ours (CAF-Score, CLAP + LALM)}} \\
\midrule
Qwen2.5-3B + MS & 0.382 & 0.351 & 0.243 & 0.485 & 0.451 & 0.316 \\
Qwen2.5-3B + LAION & 0.469 & 0.455 & 0.320 & 0.538 & 0.526 & 0.372 \\
Qwen2.5-3B + MGA & 0.504 & 0.477 & 0.334 & 0.559 & 0.545 & 0.389 \\
Qwen2.5-3B + M2D & 0.504 & 0.482 & 0.340 & 0.556 & 0.543 & 0.388 \\
Qwen2.5-7B + MS & 0.348 & 0.312 & 0.216 & 0.437 & 0.400 & 0.278 \\
Qwen2.5-7B + LAION & 0.455 & 0.435 & 0.304 & 0.519 & 0.506 & 0.356 \\
Qwen2.5-7B + MGA & 0.494 & 0.459 & 0.321 & 0.570 & 0.551 & 0.392 \\
Qwen2.5-7B + M2D & \underline{0.540} & 0.517 & \underline{0.367} & \textbf{0.609} & \underline{0.598} & \underline{0.430} \\
AF3 + MS & 0.270 & 0.234 & 0.160 & 0.362 & 0.324 & 0.223 \\
AF3 + LAION & 0.413 & 0.393 & 0.273 & 0.472 & 0.459 & 0.319 \\
AF3 + MGA & 0.452 & 0.416 & 0.289 & 0.534 & 0.522 & 0.368 \\
AF3 + M2D & 0.491 & 0.464 & 0.327 & 0.544 & 0.543 & 0.383 \\
Qwen3 + MS & 0.372 & 0.329 & 0.227 & 0.521 & 0.483 & 0.340 \\
Qwen3 + LAION & 0.466 & 0.439 & 0.306 & 0.582 & 0.572 & 0.407 \\
Qwen3 + MGA & 0.500 & 0.460 & 0.321 & 0.596 & 0.579 & 0.412 \\
Qwen3 + M2D & 0.526 & \underline{0.520} & \underline{0.367} & \underline{0.608} & \textbf{0.606} & \textbf{0.436} \\
\bottomrule
\end{tabular}%
\end{table*}

\subsubsection{BRACE-Hallucination}
\label{sec:brace_hallucination}

As shown in Table~\ref{tab:main_table}, the trends from BRACE-Main persist: S-CLAPScore outperforms CLAPScore, and FLEUR outperforms raw LALM scores. However, the highest performance here is achieved by FLEUR alone (98.18 with Qwen3-Omni-Instruct; 95.09 with AudioFlamingo3-Base) rather than CAF-Score. This does not contradict our hypothesis; rather, BRACE-Hallucination targets object-level hallucinations (e.g., changing a subject from ``man'' to ``woman''), for which the fine-grained comprehension of LALMs dominates over the contextual alignment of CLAP, whose scores remain in the 80--90 range.

We also note that this subset appears near-saturated for LALM-based evaluation in general: AQAScore likewise reports above 98 accuracy on most hallucination cases~\citep{kuan2026aqascore}, in line with our FLEUR results, indicating that object-level noun substitutions are reliably detectable by strong LALMs regardless of the surrounding framework.

Even so, CAF-Score still surpasses standalone CLAP, confirming the value of LALM calibration for hallucination detection, and outperforms the reference-based metrics: 97.11 versus CLAIR-A (91.51) and FENSE (96.40). The combination of coarse-grained CLAP alignment and fine-grained LALM evaluation is thus particularly effective at detecting specific acoustic hallucinations, offering a scalable alternative to reference-based evaluation that eliminates the dependency on costly ground-truth annotations.

\subsection{Text-to-Audio Generation Evaluation}
\label{sec:tta}

To validate the opposite direction of the alignment problem, we evaluate CAF-Score on the TTA generation task, in which the text prompt is given and the generated audio is assessed. We conduct experiments on the \textbf{RELATE}~\citep{kanamori2025relate} and \textbf{PAM}~\citep{deshmukh2024pam} benchmarks, which provide gold-standard human subjective ratings for TTA generation quality, and compare against \textbf{AQAScore}~\citep{kuan2026aqascore}.

\subsubsection{RELATE}
On the RELATE dataset (Table~\ref{tab:tta_full_results}), AQAScore with Qwen-2.5-Omni-7B remains the strongest single configuration. We attribute this to its verification formulation, which asks directly whether the audio contains the sound events described by the text---posed in exactly the direction the TTA task requires---whereas CAF-Score uses a generic grading prompt shared, up to perspective inversion, across both alignment directions. Nevertheless, under the same backbones, CAF-Score improves upon standalone LALM (Raw Score, FLEUR) and standalone CLAP metrics in most configurations---the exceptions almost all involve MS-CLAP, by far the weakest CLAP backbone on this benchmark---and the best CAF-Score configuration (Qwen-2.5-Omni-7B + M2D-CLAP) is competitive with AQAScore. This reinforces the hypothesis that LALM-based calibration effectively refines coarse-grained semantic alignment.

\subsubsection{PAM}
On the PAM benchmark, CAF-Score configurations utilizing the Qwen family---including both Qwen3-Omni-Instruct and Qwen-2.5-Omni-7B---combined with M2D-CLAP achieve the best overall correlations, surpassing AQAScore across all three correlation measures. Across various model scales, CAF-Score improves upon standalone metrics in most configurations.

\paragraph{Generalizability.}
Taken together, the TTA results demonstrate that CAF-Score, configured entirely on BRACE, generalizes to a distinct alignment task, distinct data distributions (generated rather than natural audio), and a distinct evaluation protocol (correlation with graded ratings rather than pairwise preference accuracy). The comparison with AQAScore across both directions illustrates the underlying design trade-off. AQAScore is the strongest metric on RELATE, yet on the captioning direction it reports accuracy comparable to CLAPScore and identifies its inconsistency in text-selection settings as a limitation~\citep{kuan2026aqascore}. In contrast, a single CAF-Score configuration attains the best reference-free accuracy on BRACE, the best correlations on PAM, and competitive performance on RELATE. To our knowledge, no other reference-free metric is simultaneously competitive in both alignment directions without per-task adaptation, supporting our central claim that hybridizing contrastive and generative paradigms yields a reliable, direction-agnostic assessment of audio-text alignment.

\subsection{Metric Design Analysis}
\label{sec:design_analysis}
Analyses validating the three architectural choices underlying CAF-Score are provided in Appendix~\ref{sec:appendix_design}. In brief: (i) on variable-length audio, sliding windows outperform truncation, while Max and average pooling perform comparably; we adopt Max pooling to preserve brief but salient events; (ii) the hybrid weighting ($0<\alpha<1$) consistently outperforms either component alone, and an entropy-based sample-level adaptive $\alpha$ did not improve over the fixed value; and (iii) FLEUR's advantage over raw LALM scores stems from resolving the frequent ties of discrete outputs, whose rates exceeded 50\% for AudioFlamingo3.

\section{Conclusion}
We proposed the \textbf{CAF-Score}, a reference-free metric for audio-text alignment evaluation that calibrates the coarse-grained semantic alignment of CLAP with the fine-grained semantic comprehension and syntactic awareness of LALMs. Across three human-annotated benchmarks spanning audio captioning and TTA generation evaluation, a single configuration---one shared prompt formulation and one fusion weight transferred without re-tuning---achieves strong agreement with human judgment. It also remains competitive with reference-based baselines, including in the challenging human-human comparison setting and in hallucination detection. Our results consistently show that the hybrid approach ($0 < \alpha < 1$) outperforms standalone CLAP in nearly all settings and standalone LALM evaluation in most; the exception is object-level hallucination detection, where LALM comprehension alone suffices. This complementarity is visible in the qualitative analysis (Appendix~\ref{sec:qualitative}): CLAP provides stable acoustic grounding when the LALM struggles with linguistically similar captions, while the LALM corrects CLAP's superficial keyword-matching errors. Integrating contrastive and generative paradigms is thus a promising direction for future reference-free metrics, and CAF-Score offers a robust, direction-agnostic evaluation framework for the rapidly evolving field of audio-language modeling.

\section*{Limitations}
Our framework has several limitations. First, incorporating LALMs introduces substantial computational overhead compared to lightweight embedding-based models such as CLAP; although practical for offline evaluation, CAF-Score trades computational efficiency for evaluation accuracy, which may limit its use in large-scale or latency-sensitive settings such as online monitoring or training-time reward computation.

Second, CAF-Score inherits the failure modes of its underlying components. When both CLAP and the LALM simultaneously misalign with human judgment---the ``double failure'' cases analyzed in Appendix~\ref{sec:appendix_failure}---the metric cannot recover. Relatedly, the fixed weighting parameter $\alpha=0.8$ may not be universally optimal across all audio-text pairs; as discussed in Appendix~\ref{sec:appendix_alpha}, entropy-based adaptive weighting did not outperform the fixed configuration on average, but exploring alternative adaptive strategies remains a promising direction for future work.

Third, our generic single-prompt formulation, adapted directly from the vision domain, deliberately trades task specialization for cross-direction robustness: on RELATE, the verification-style formulation of AQAScore outperforms CAF-Score, while CAF-Score prevails on PAM and in the captioning direction, where AQAScore reports accuracy comparable to CLAPScore~\citep{kuan2026aqascore}. We view this as an inherent trade-off between the two design philosophies rather than a deficiency of either; combining CLAP calibration with verification-style or otherwise task-adapted prompting---ideally without sacrificing direction-agnostic applicability---is an interesting avenue for future work.

Finally, our evaluation covers general (environmental) audio in English; extending the framework to music, speech-centric content, and other languages remains future work. While we validate on three human-annotated benchmarks across two tasks, specialized reference-free evaluation resources in the audio-language domain remain limited, and broader validation would further strengthen confidence in the metric. We hope this work encourages the development of more diverse evaluation resources for reference-free audio-text alignment.

\section*{Acknowledgments}
This work was supported by the Institute of Information \& Communications Technology Planning \& Evaluation (IITP) grant funded by the Korea government (MSIT) (No.~RS-2025-25441313, Professional AI Talent Development Program for Multimodal AI Agents, 50\%), by the Korea Planning \& Evaluation Institute of Industrial Technology (KEIT) grant funded by the Korea government (MOTIR) (No.~RS-2026-25625718, Development and Demonstration of a Lightweight Manufacturing Foundation Model Specialized for Complex Assembly Processes, 25\%), and by the Institute of Information \& Communications Technology Planning \& Evaluation (IITP) under the Artificial Intelligence Innovation Human Resources Development (IITP-RS-2026-25547954) grant funded by the Korea government (MSIT) (25\%).

\bibliography{mybib}

\appendix

\section{Experimental Setup Details}
\label{sec:appendix_setup}

\subsection{BRACE Details and Evaluation Protocol}
\label{sec:appendix_brace}

The audio durations in BRACE typically range from \SIrange{5}{30}{s}, with AudioCaps clips specifically limited to a maximum of \SI{10}{s}, and caption lengths are generally under 20 words.

The original BRACE benchmark evaluates LALMs using a pairwise comparison protocol, in which two captions are presented simultaneously and the model selects the better one. While effective for eliciting preferences, pairwise prompting can encourage relative judgments based on superficial cues---such as fluency, length, or lexical priors---rather than on explicit audio grounding. This makes it difficult to attribute decisions solely to audio-text alignment. To enable a more grounded and analyzable evaluation, we adopt a single-caption evaluation protocol, in which each caption is scored independently against the corresponding audio by both CLAP and the LALM, yielding S-CLAPScore and FLEUR for each audio-caption pair. This approach allows fine-grained analysis at the individual caption level, while still permitting reconstruction of the original BRACE pairwise preferences by comparing the independently obtained scores. We use the original BRACE data without modification, changing only the evaluation protocol.

\subsection{TTA Evaluation Setup}
\label{sec:appendix_tta_setup}

For the TTA benchmarks (RELATE and PAM), we follow the methodology established in Section~\ref{sec:method}, utilizing the CLAPScore as the foundation for coarse-grained alignment. Given that TTA samples in these benchmarks typically range around 10 seconds, the impact of sliding windows was found to be minimal. To adapt the evaluation for the TTA task, we modified the LALM evaluation prompt by simply inverting the perspective between the audio and the caption; the exact prompt template is provided in Appendix~\ref{sec:appendix_prompts}. The fusion weight $\alpha=0.8$ selected on BRACE-Main was applied without any re-tuning.

\subsection{Baseline Models}
\label{sec:appendix_baselines}

To ensure a comprehensive evaluation, we selected representative models spanning diverse architectures and training paradigms.

\noindent\textbf{CLAP Models.} We used four representative variants to assess audio-text alignment:
\begin{itemize}
    \item \textbf{MS-CLAP}\footnote{\url{https://github.com/microsoft/CLAP}}: The 2023 released version.
    \item \textbf{LAION-CLAP}\footnote{\url{https://huggingface.co/laion/clap-htsat-unfused}}: The \texttt{`laion/clap-htsat-unfused'} checkpoint from Hugging Face.
    \item \textbf{MGA-CLAP}\footnote{\url{https://github.com/Ming-er/MGA-CLAP}}: Pre-trained weights downloaded via the official Google Drive link provided in the repository.
    \item \textbf{M2D-CLAP}\footnote{\url{https://github.com/nttcslab/m2d}}: Pre-trained weights provided in the repository.
\end{itemize}

\noindent\textbf{LALMs.} We evaluated two prominent families, considering both standard and reasoning variants:
\begin{itemize}
    \item \textbf{AudioFlamingo3}\footnote{\url{https://huggingface.co/nvidia/audio-flamingo-3-hf}}: We evaluated both the \textit{Base} and \textit{Think} variants.
    \item \textbf{Qwen3-Omni}: We evaluated both the \textit{Instruct}\footnote{\url{https://huggingface.co/Qwen/Qwen3-Omni-30B-A3B-Instruct}} and \textit{Thinking}\footnote{\url{https://huggingface.co/Qwen/Qwen3-Omni-30B-A3B-Thinking}} variants.
\end{itemize}
On the TTA benchmarks, we additionally evaluate the \textbf{Qwen-2.5-Omni}~\citep{wei2025qwen25omni} family (3B\footnote{\url{https://huggingface.co/Qwen/Qwen2.5-Omni-3B}} and 7B\footnote{\url{https://huggingface.co/Qwen/Qwen2.5-Omni-7B}}) as backbones. This enables a backbone-controlled comparison with AQAScore~\citep{kuan2026aqascore}, which is built on Qwen-2.5-Omni: evaluating CAF-Score on the same backbones isolates the contribution of the framework from that of the underlying model.

\subsection{Implementation Details}
\label{sec:appendix_implementation}
All experiments were conducted on a single NVIDIA A100 (80GB) GPU. For the Qwen3-Omni family, we employed the vLLM library on the same hardware to efficiently handle the large model size, with the GPU memory utilization parameter (`gpu\_memory\_utilization') set to 0.95. To ensure deterministic evaluation across all models, the temperature was set to 0.0 for all LALM generations.

\subsection{Evaluation Protocol}
\label{sec:evaluation_protocol}
Our experiments were designed to validate each component of the CAF-Score:

\begin{itemize}
    \item \textbf{Sliding-window Analysis (CLAP):} We compared the standard CLAPScore with S-CLAPScore. For CLAPScore, to ensure fair comparison and reproducibility, the input audio was truncated to the model's maximum supported duration: \SI{7}{s} for MS-CLAP and \SI{10}{s} for the other CLAP variants. Audio shorter than the window length is zero-padded to meet the encoders' input requirements.

    \item \textbf{Scoring Method Analysis (LALM):} To evaluate the robustness of FLEUR and the alignment capability of LALMs, we compared the probabilistic FLEUR metric against raw discrete generation scores. This analysis demonstrates the necessity of probabilistic smoothing (using token log-probabilities) for reliable evaluation.

    \item \textbf{CAF-Score Configuration:} For the final CAF-Score calculation, we exclusively used the non-reasoning variants of LALMs (AudioFlamingo3-Base, Qwen3-Omni-Instruct), as reasoning variants exhibited lower stability in score generation, as discussed in Section~\ref{sec:brace_main}.
\end{itemize}

\subsection{Sources of Reported Results}
\label{sec:reproducibility}
All results on the BRACE benchmark in Table~\ref{tab:main_table}---including the reference-based baselines FENSE and CLAIR-A---were reproduced by us under the single-caption evaluation protocol described in Appendix~\ref{sec:appendix_brace}, ensuring that all metrics are compared on identical reconstructed pairwise preferences. For the TTA benchmarks (Table~\ref{tab:tta_full_results}), the RELATE and AQAScore rows are taken from their respective original papers~\citep{kanamori2025relate,kuan2026aqascore}, as the AQAScore implementation was not publicly available at the time of writing; all remaining results (CLAPScore, Raw Score, FLEUR, and CAF-Score across all backbones) were obtained by our own evaluation under the setup described in Appendix~\ref{sec:appendix_tta_setup}.

\section{Metric Design Analysis}
\label{sec:appendix_design}

This appendix validates the three architectural choices underlying CAF-Score: the sliding-window aggregation strategy on the CLAP branch, the fixed fusion weight $\alpha$, and the probabilistic FLEUR scoring on the LALM branch.

\subsection{Sliding Windows and Pooling}
\label{sec:appendix_pooling}
On Clotho, whose audio is variable-length, sliding windows with pooling outperform the non-pooling baseline across most CLAP backbones (e.g., MS-CLAP improves from 71.65 to 73.21 with Max pooling; Table~\ref{tab:ablation_pooling}). Between the two pooling strategies, Max pooling performs best or comparably across backbones, and we adopt it as the default, as selecting the most salient segment mitigates the dilution of brief but critical events under averaging.

\begin{table}[t]
\centering
\caption{Comparison of pooling strategies for CAF-Score (using Qwen3-Omni-Instruct) evaluated on Clotho-Main across different CLAP backbones. The pooling methods are: no pooling (X), average pooling (Avg), and maximum pooling (Max).}
\label{tab:ablation_pooling}
\setlength{\tabcolsep}{4pt}
\footnotesize
\begin{tabular}{llcccc}
\toprule
\multirow{2}{*}{\textbf{Backbone}} & \multirow{2}{*}{\textbf{Pool}} & \multicolumn{4}{c}{\textbf{Clotho-Main (CAF-Score)}} \\
\cmidrule(lr){3-6}
 & & HH & HM & MM & \textbf{Total} \\
\midrule
\multirow{3}{*}{MS-CLAP} & X & \textbf{68.86} & 78.00 & 67.56 & 71.65 \\
 & Avg & 66.47 & 78.59 & 69.48 & 72.54 \\
 & Max & 67.66 & \textbf{79.17} & \textbf{70.07} & \textbf{73.21} \\
\midrule
\multirow{3}{*}{LAION-CLAP} & X & 61.08 & 78.59 & 67.85 & 71.06 \\
 & Avg & \textbf{61.68} & \textbf{80.35} & 67.70 & 71.72 \\
 & Max & \textbf{61.68} & 80.16 & \textbf{68.00} & \textbf{71.80} \\
\midrule
\multirow{3}{*}{MGA-CLAP} & X & 71.26 & 74.46 & 64.74 & 69.21 \\
 & Avg & 70.66 & \textbf{75.83} & \textbf{65.04} & \textbf{69.80} \\
 & Max & \textbf{73.65} & 75.44 & 63.85 & 69.43 \\
\midrule
\multirow{3}{*}{M2D-CLAP} & X & \textbf{73.05} & 77.01 & 66.22 & 71.13 \\
 & Avg & 70.06 & 77.01 & \textbf{66.81} & 71.06 \\
 & Max & 70.66 & \textbf{77.80} & 66.52 & \textbf{71.28} \\
\bottomrule
\end{tabular}
\end{table}

\subsection{Weighting Parameter $\alpha$}
\label{sec:appendix_alpha}

To determine the optimal balance between CLAP and LALM contributions, we varied the weighting parameter $\alpha$, combining Qwen3-Omni-Instruct with different CLAP backbones. The results of this experiment are shown in Figure~\ref{fig:alpha}.

We observed that for both LAION-CLAP and M2D-CLAP, performance peaked at $\mathbf{\alpha=0.8}$. In contrast, MS-CLAP and MGA-CLAP favored a higher contribution from the LALM component, with peak performance at $\alpha=0.2$ and $\alpha=0.5$, respectively. Overall, the highest scores were achieved by the combinations using M2D-CLAP + Qwen3-Omni-Instruct (73.11) and LAION-CLAP + Qwen3-Omni-Instruct (72.24). These results show that the hybrid approach ($0 < \alpha < 1$) consistently outperforms either CLAP ($\alpha=1$) or LALM ($\alpha=0$) alone. We emphasize that $\alpha=0.8$ was selected once on BRACE-Main based on the average performance across backbones and applied unchanged to all other evaluations; the TTA results in Section~\ref{sec:tta} were thus obtained held-out, mitigating the risk of benchmark-specific calibration.

\paragraph{Sample-level adaptive weighting.}
We further investigated whether a sample-level adaptive weighting could improve upon the fixed $\alpha$. Specifically, we set $\text{adaptive}\,\alpha$ to the normalized entropy of the digit-token probability distribution at the first decimal position used in the FLEUR calculation:
\begin{equation}
\text{adaptive}\,\alpha = \frac{H(\tilde{p})}{H_{\max}},\ \
    H(\tilde{p}) = -{\textstyle\sum_{i}}\, \tilde{p}(i) \log \tilde{p}(i)
\end{equation}
where $\tilde{p}$ denotes the digit-token probabilities renormalized to sum to one over the digit tokens (since only the top-20 log-probabilities are available for the Qwen-family models at inference time; Appendix~\ref{sec:appendix_fleur_example}), and $H_{\max}$ is the maximum entropy of the distribution, so that $\text{adaptive}\,\alpha \in [0,1]$. The rationale is that higher entropy indicates greater LALM uncertainty, warranting a larger weight on S-CLAPScore to maintain evaluation stability. However, as shown in Table~\ref{tab:adaptive}, the optimal fixed $\alpha$ consistently outperformed the adaptive approach across all CLAP backbones.

We attribute this result primarily to an information redundancy between the adaptive weight and FLEUR itself. Because FLEUR is computed as an expectation over the digit-token distribution, a diffuse (high-entropy) distribution already pulls the score away from extreme values rather than committing to a single digit; the LALM's uncertainty is thus partially encoded in the FLEUR score before any reweighting is applied. Consequently, entropy-based modulation of $\alpha$ injects little information beyond what the smoothed score already carries, while introducing sample-level variability in the fusion weight. In practice, the digit-token distributions were often diffuse enough that the adaptive scheme systematically shifted weight toward S-CLAPScore, suppressing the fine-grained FLEUR signal that motivates the hybrid design. A fixed weighting therefore provides more robust performance than sample-level adaptation.

\paragraph{Score scales and normalization.}
Although S-CLAPScore (cosine similarity) and FLEUR lie on comparable numeric scales, their empirical distributions concentrate in different regions---S-CLAPScore in a narrow low-value band and FLEUR toward the upper end of $[0,1]$---which raises the question of whether the two components should be normalized before fusion. We note that explicit normalization is unnecessary under our evaluation protocols: any per-component affine rescaling followed by a weighted combination differs from the unnormalized combination only by a positive affine transformation, for some value of $\alpha$. Since pairwise preference accuracy, SRCC, KTAU, and Pearson correlation are all invariant to positive affine transformations, normalized fusion is order-equivalent to unnormalized fusion with a different weight, and the $\alpha$ sweep in Figure~\ref{fig:alpha} therefore already covers all per-component affine normalization schemes. Accordingly, $\alpha$ should be interpreted as jointly absorbing the scale discrepancy and the desired balance between the two components, rather than as a direct measure of relative trust between them; avoiding normalization also removes the need to estimate normalization statistics from data, preserving the tuning-free design of CAF-Score.

\begin{figure}[t]
    \centering
    \begin{tikzpicture}
        \pgfplotsset{
            width=0.85\columnwidth,
            height=5.5cm,
            every axis plot/.append style={very thick, mark size=2.5pt},
            legend style={at={(0.98,0.02)}, anchor=south east, draw=none, fill=white, fill opacity=0.8, font=\small},
            xlabel={Weighting Parameter $\alpha$},
            ylabel={Performance Score},
            xmin=0, xmax=1,
            ymin=60, ymax=75,
            xtick={0, 0.2, 0.5, 0.8, 1.0},
            xticklabels={0\\(FLEUR), 0.2, 0.5, 0.8, 1\\($S_{\text{S-CLAP}}$)},
            x tick label style={font=\scriptsize, align=center},
            y tick label style={font=\small},
            grid=major,
            grid style={dashed, gray!30}
        }

        \definecolor{colorMS}{RGB}{215,48,39}    
        \definecolor{colorLAION}{RGB}{252,141,89} 
        \definecolor{colorMGA}{RGB}{145,191,219}  
        \definecolor{colorM2D}{RGB}{69,117,180}   

        \begin{axis}
            \addplot[color=colorMS, mark=square*] coordinates {
                (0, 68.31) (0.2, 70.11) (0.5, 69.55) (0.8, 68.39) (1, 61.30)
            };
            \addlegendentry{MS-CLAP}

            \addplot[color=colorLAION, mark=triangle*] coordinates {
                (0, 68.31) (0.2, 70.75) (0.5, 71.99) (0.8, 72.24) (1, 69.59)
            };
            \addlegendentry{LAION-CLAP}

            \addplot[color=colorMGA, mark=diamond*] coordinates {
                (0, 68.31) (0.2, 70.35) (0.5, 71.32) (0.8, 70.79) (1, 67.06)
            };
            \addlegendentry{MGA-CLAP}

            \addplot[color=colorM2D, mark=*] coordinates {
                (0, 68.31) (0.2, 69.87) (0.5, 71.28) (0.8, 73.11) (1, 68.55)
            };
            \addlegendentry{M2D-CLAP}

        \end{axis}
    \end{tikzpicture}
    \caption{Performance variation of CAF-Score across different weighting parameters $\alpha$ on BRACE-Main.}
    \label{fig:alpha}
\end{figure}

\begin{table}[t]
    \caption{Comparison of optimal fixed $\alpha$ versus entropy-based $\text{adaptive}\,\alpha$ on BRACE-Main.}
    \label{tab:adaptive}
    \centering
    \footnotesize
    \resizebox{\columnwidth}{!}{%
    \begin{tabular}{lcccc}
        \toprule
        Method & MS-CLAP & LAION-CLAP & MGA-CLAP & M2D-CLAP \\
        \midrule
        Best Fixed $\alpha$ & 0.2 & 0.8 & 0.5 & 0.8 \\
        \midrule
        Fixed $\alpha$ & \textbf{70.11} & \textbf{72.24} & \textbf{71.32} & \textbf{73.11} \\
        Adaptive $\alpha$ & 70.03 & 71.99 & 71.15 & 69.99 \\
        \bottomrule
    \end{tabular}
    }
\end{table}

\subsection{Raw Scores versus FLEUR}
\label{sec:appendix_tie}
\begin{figure}[t]
    \centering
    \includegraphics[width=\columnwidth]{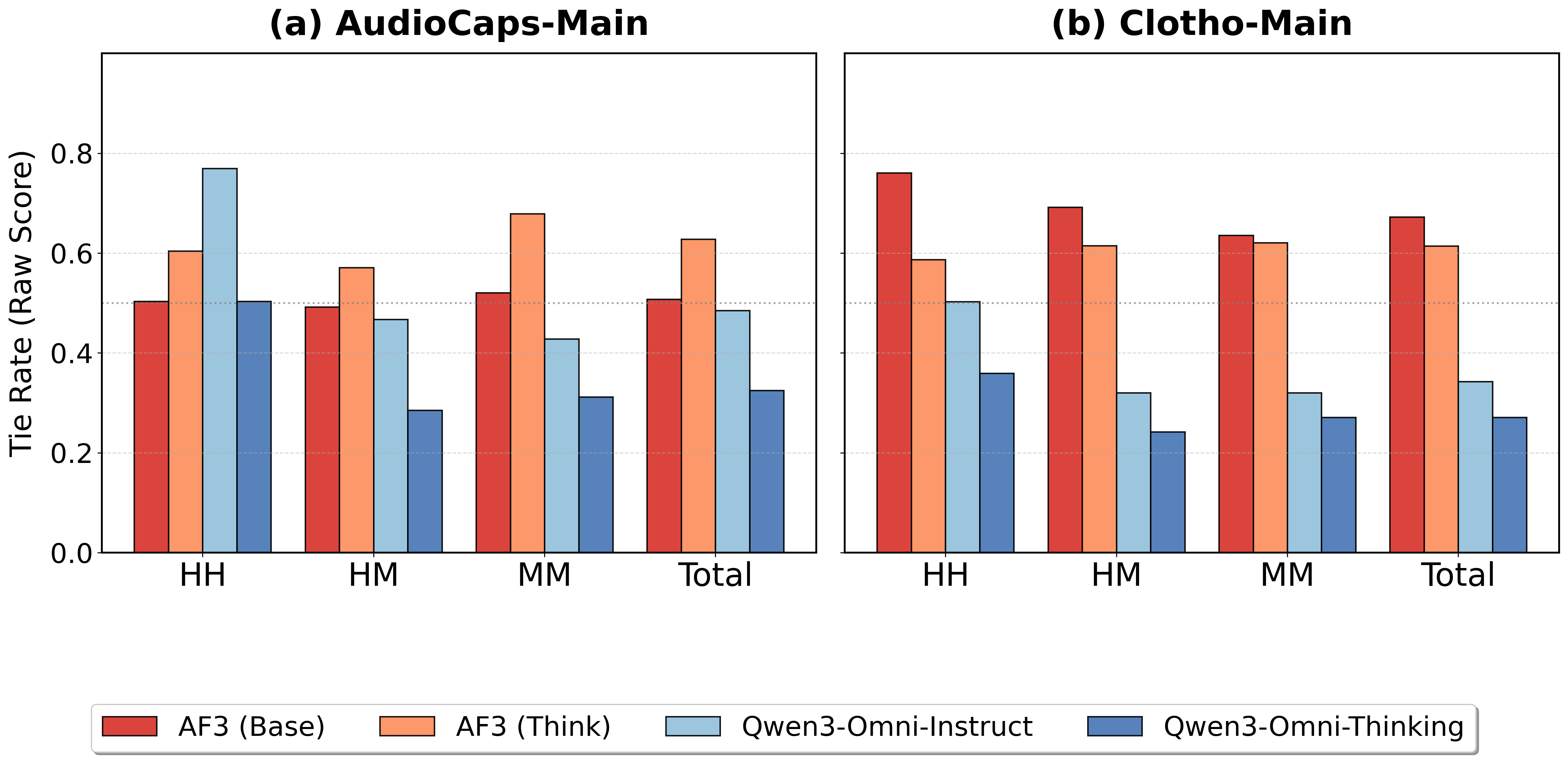}
    \caption{Tie rates in raw scores for AudioFlamingo3 (AF3) and Qwen3-Omni models.}
    \label{fig:tie_rate}
\end{figure}

FLEUR's advantage over raw LALM scores (Table~\ref{tab:main_table}) stems from ties: raw outputs collapse onto a few discrete values---AudioFlamingo3 assigned ``0.0'' or ``1.0'' in over 80\% of tie cases, and Qwen3-Omni behaved similarly---a documented issue in LLM-based evaluation~\citep{wu2024clair, lee2024fleur}. Tie rates exceeded 50\% for AudioFlamingo3 and remained substantial even for Qwen3-Omni's reasoning variant (Figure~\ref{fig:tie_rate}), making probabilistic smoothing via FLEUR the more reliable approach.

\section{FLEUR Calculation Example}
\label{sec:appendix_fleur_example}

For illustrative purposes, consider a simplified example where the model outputs ``0.85''. Instead of taking this value directly, FLEUR examines the token-level probabilities at each decimal position. Suppose the first decimal place has $P(\text{`8'})=0.7$ and $P(\text{`9'})=0.3$, and the second decimal place has $P(\text{`5'})=0.6$ and $P(\text{`4'})=0.4$, with all other digit probabilities near zero. The FLEUR score then becomes:
\begin{equation}
\begin{split}
    \text{FLEUR} &= 0.1 \times (8 \times 0.7 + 9 \times 0.3) \\
    &\quad + 0.01 \times (5 \times 0.6 + 4 \times 0.4) \\
    &= 0.1 \times 8.3 + 0.01 \times 4.6 \\
    &= 0.876
\end{split}
\end{equation}

In practice, only the top-20 log-probabilities are available for the Qwen-family models, so digit tokens outside the top 20 are assigned zero probability; for AudioFlamingo3, the full distribution is used. When the raw output is ``1.0'’, the score is computed at the integer-digit position as $1.0 \cdot p(\text{`1'}) + 0.9 \cdot p(\text{`0'})$. Outputs that cannot be parsed as a number in $[0, 1]$ receive a FLEUR score of 0.

\section{Evaluation Prompts}
\label{sec:appendix_prompts}

The exact prompt templates used for the LALM-based FLEUR evaluation are shown below. The audio captioning prompt is adapted from~\citet{lee2024fleur} by replacing ``image'' with ``audio''; the TTA prompt inverts the perspective between the audio and the caption.

\begin{tcolorbox}[colback=gray!5, colframe=gray!50, title=Evaluation Prompt (Audio Captioning)]
\footnotesize\ttfamily
Your task is to evaluate and rate the caption on a scale of 0.0 to 1.0 based on the given Grading Criteria. (Print Real Number Score ONLY)

\vspace{0.2cm}
Grading Criteria:

0.0: The caption does not describe the audio at all.\\
1.0: The caption accurately and clearly describes the audio.

\vspace{0.2cm}
Caption: \{pred\_caption\}

\vspace{0.2cm}
Score(Choose a rating from 0.0 to 1.0):
\end{tcolorbox}

\begin{tcolorbox}[colback=gray!5, colframe=gray!50, title=Evaluation Prompt (TTA)]
\footnotesize\ttfamily
Your task is to evaluate and rate the audio on a scale of 0.0 to 1.0 based on the given Grading Criteria. (Print Real Number Score ONLY)

\vspace{0.15cm}
Grading Criteria:

0.0: The audio does not describe the caption at all.\\
1.0: The audio accurately and clearly describes the caption.

\vspace{0.15cm}
Caption: \{caption\}

\vspace{0.15cm}
Score (Choose a rating from 0.0 to 1.0):
\end{tcolorbox}

\section{Qualitative Analysis}
\label{sec:qualitative}

We present qualitative examples of CAF-Score to illustrate its strengths and limitations.

\subsection{Success Cases}
CAF-Score effectively combines the strengths of both models to mitigate their individual limitations. We categorize these success cases into two complementary scenarios.

\begin{figure}[h]
    \centering
    \includegraphics[width=0.8\linewidth]{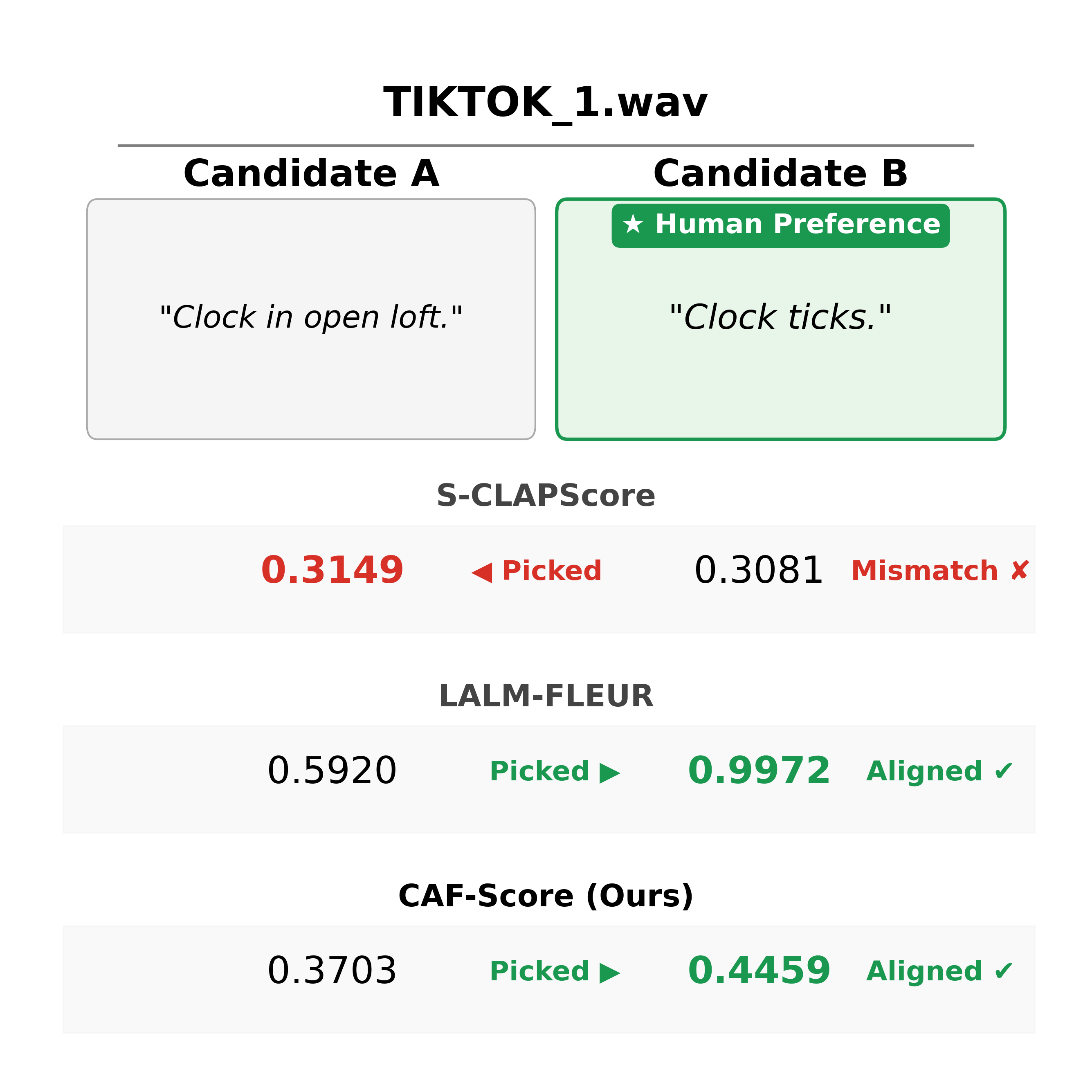}
    \caption{Success case: LALM-driven Correction on \texttt{TIKTOK\_1.wav}.}
    \label{fig:case_study_a}
\end{figure}

\noindent\textbf{LALM-driven Correction:}
In Figure~\ref{fig:case_study_a}, S-CLAPScore assigns a higher score to an unpreferred caption than to the preferred one, due to superficial keyword overlap between ``Clock'' and the acoustic event ``ticks''. CAF-Score, however, leverages the fine-grained understanding of the LALM to detect this mismatch, adjusting the final score to align with human preference.

\begin{figure}[h]
    \centering
    \includegraphics[width=0.8\linewidth]{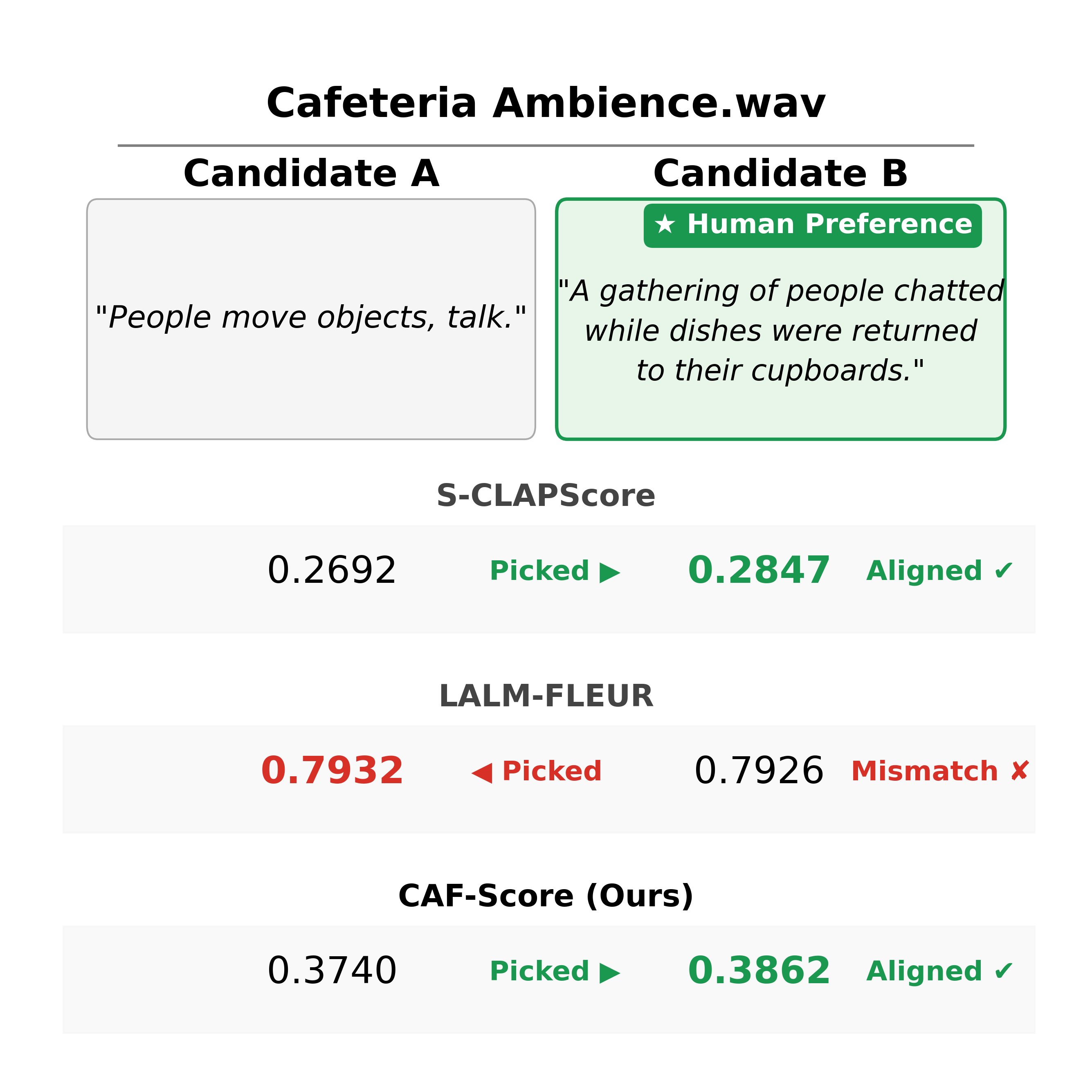}
    \caption{Success case: CLAP-driven Stability on \texttt{Cafeteria Ambience.wav}.}
    \label{fig:case_study_b}
\end{figure}

\noindent\textbf{CLAP-driven Stability:}
Conversely, in Figure~\ref{fig:case_study_b}, both caption candidates are linguistically valid, causing the LALM to show weak and ambiguous discrimination. In this case, CAF-Score relies on the robust coarse-grained alignment of S-CLAPScore to correctly identify the more accurate caption. Together, these examples highlight the complementary nature of our hybrid approach: the LALM compensates for CLAP's shallow matching, while CLAP stabilizes the LALM's uncertain judgments.

\subsection{Failure Cases}
\label{sec:appendix_failure}

Since the performance of CAF-Score is inherently bounded by the capabilities of the underlying models, we also analyzed instances where CAF-Score failed to align with human preference. We identify three failure types, each stemming from a different mode of model dependency, with an illustrative example for each.

\begin{figure}[h]
    \centering
    \includegraphics[width=0.75\linewidth]{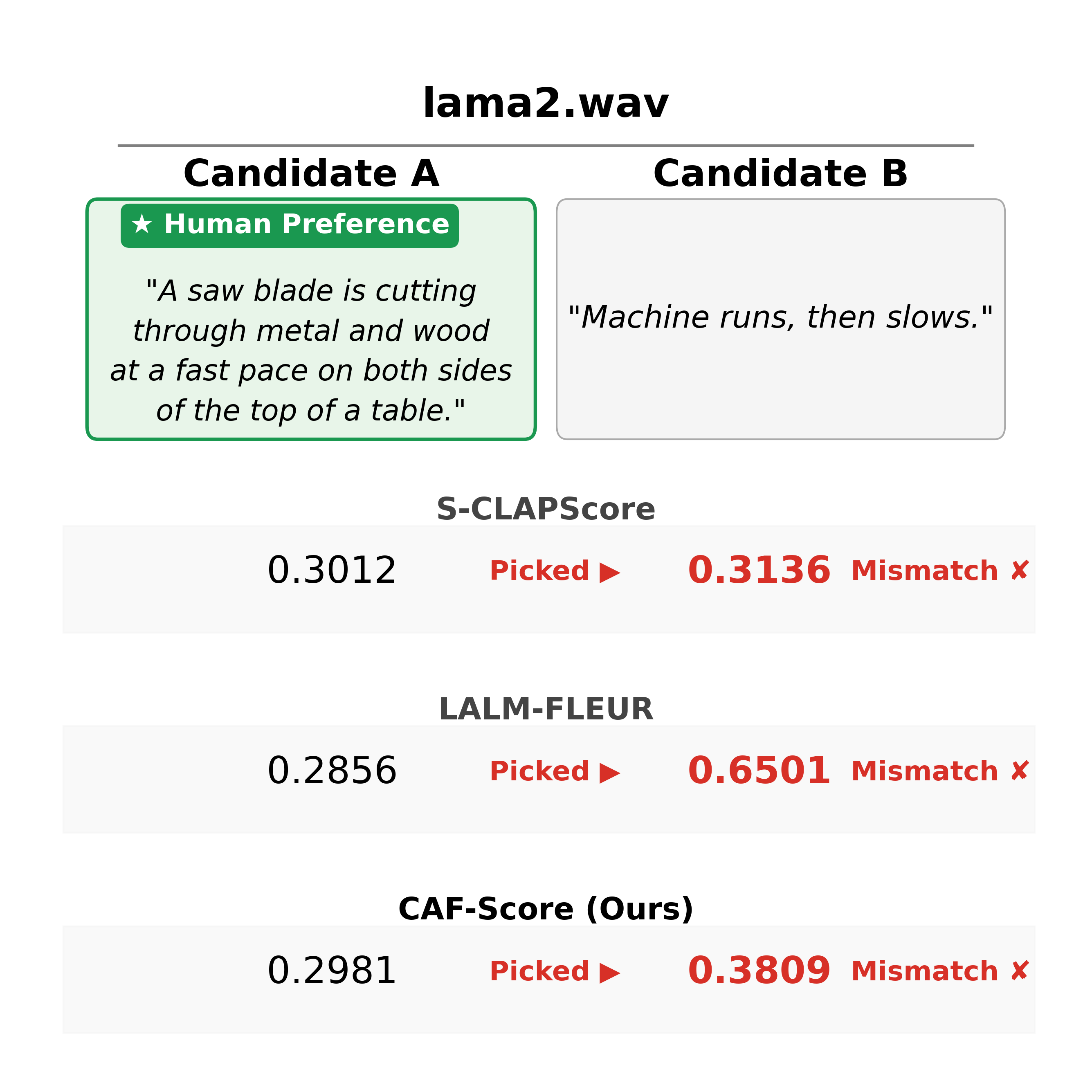}
    \caption{Double Failure on \texttt{lama2.wav}.}
    \label{fig:fail_type1}
\end{figure}

\noindent\textbf{Double Failure:}
As shown in Figure~\ref{fig:fail_type1}, when both CLAP and the LALM are simultaneously misaligned with human preference, CAF-Score is unable to rectify the prediction. This typically occurs when the audio contains ambiguous sounds that mislead both models at the same time.

\begin{figure}[h]
    \centering
    \includegraphics[width=0.8\linewidth]{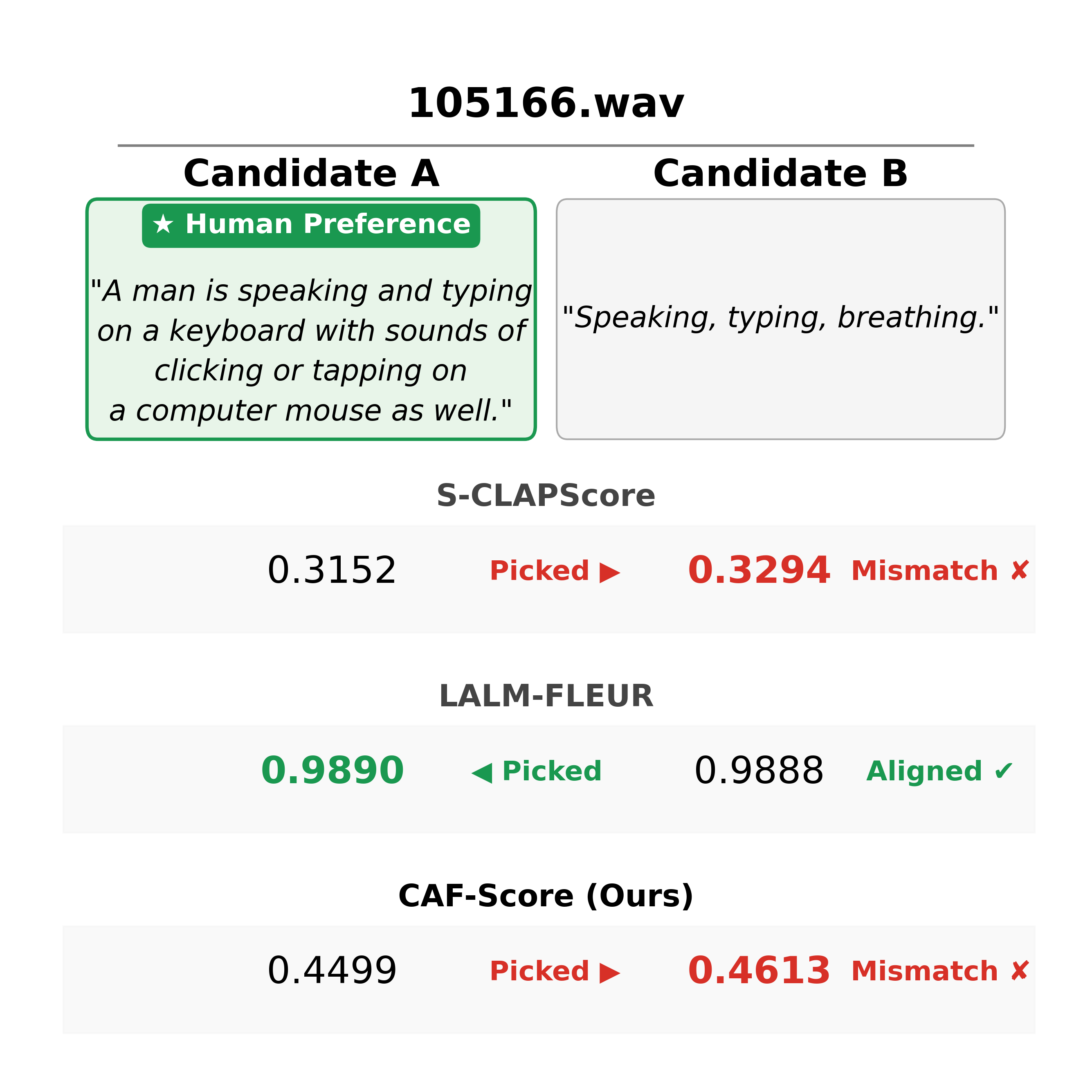}
    \caption{Insufficient Calibration on \texttt{105166.wav}.}
    \label{fig:fail_type2}
\end{figure}

\noindent\textbf{Insufficient Calibration:}
In some cases, CLAP favors the incorrect caption while the LALM correctly prefers the human-selected one. However, as illustrated in Figure~\ref{fig:fail_type2}, the LALM assigned extremely high and nearly identical scores to both candidates, producing an insufficient margin to overcome the weighted contribution of the incorrect CLAPScore.

\begin{figure}[h]
    \centering
    \includegraphics[width=0.75\linewidth]{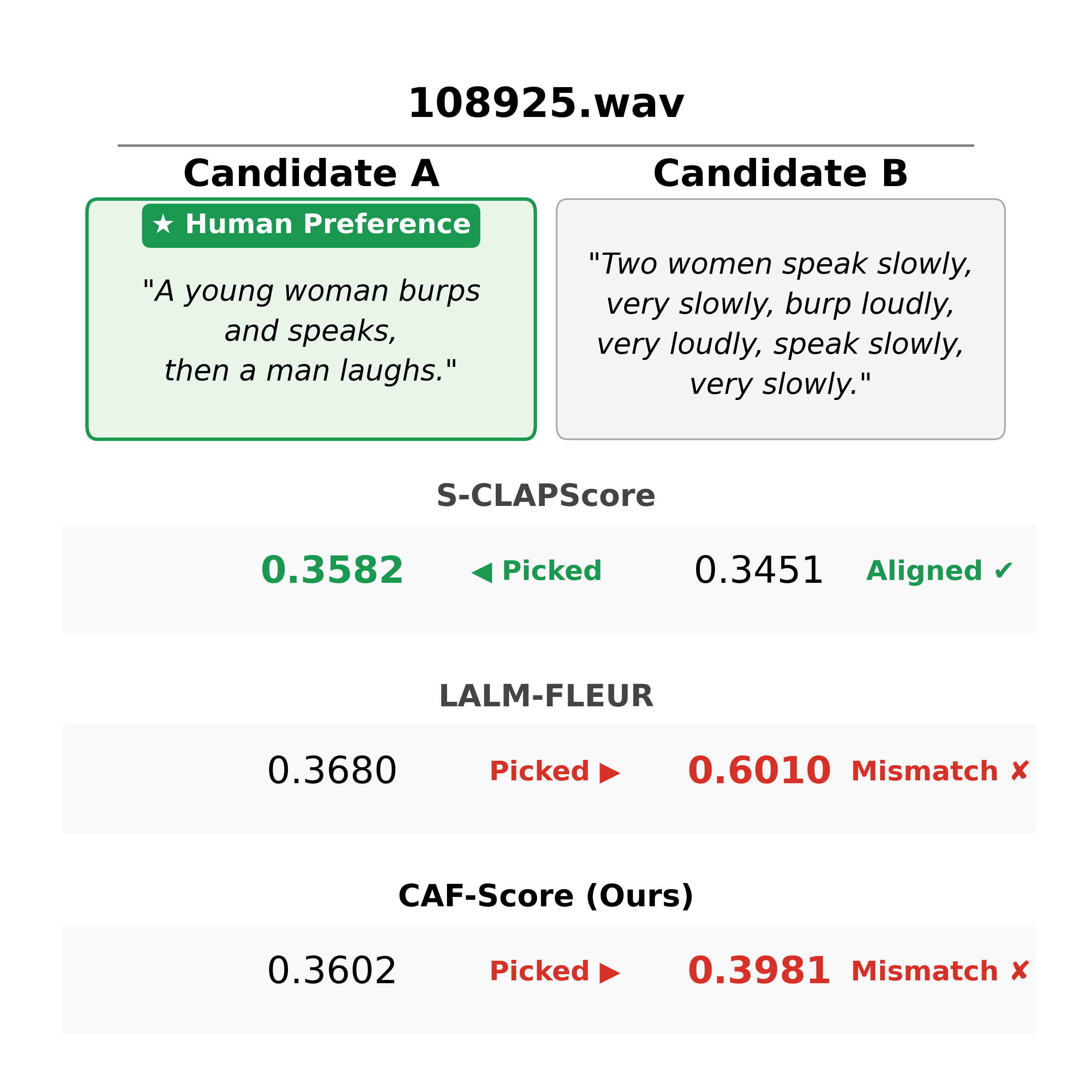}
    \caption{Over-calibration on \texttt{108925.wav}.}
    \label{fig:fail_type3}
\end{figure}

\noindent\textbf{Over-calibration:}
Conversely, CLAP may correctly identify the preferred caption, but the LALM makes an erroneous judgment due to hallucination or hypersensitivity to syntax. As shown in Figure~\ref{fig:fail_type3}, FLEUR exhibited an excessive preference for the non-preferred candidate, effectively overriding the correct alignment from CLAP and leading to a wrong prediction.

\end{document}